\documentclass[conference]{IEEEtran}

\usepackage{verbatim}
\usepackage{epsfig,amsmath,amssymb,latexsym}
\usepackage{color,epsf,psfrag,hhline,textcomp}
\usepackage{graphicx}

\usepackage[ruled]{algorithm2e}

\makeatletter
\def\ps@headings{%
\def\@oddhead{\mbox{}\scriptsize\rightmark \hfil \thepage}%
\def\@evenhead{\scriptsize\thepage \hfil \leftmark\mbox{}}%
\def\@oddfoot{}%
\def\@evenfoot{}}
\makeatother
\def\_{\rule{.3em}{.15ex}}      

 \normalmarginpar
\newcommand {\mymarginpar}[1]{\marginpar{#1}}
\renewcommand {\marginpar}[1]{}

\def\_{\rule{.3em}{.15ex}}      

\newcommand{\ls}[1]
   {\dimen0=\fontdimen6\the\font
    \lineskip=#1\dimen0
    \advance\lineskip.5\fontdimen5\the\font
    \advance\lineskip-\dimen0
    \lineskiplimit=.9\lineskip
    \baselineskip=\lineskip
    \advance\baselineskip\dimen0
    \normallineskip\lineskip
    \normallineskiplimit\lineskiplimit
    \normalbaselineskip\baselineskip
    \ignorespaces
   }

\newcommand {\bearn}{\begin{eqnarray*}}
\newcommand {\eearn}{\end{eqnarray*}}
\newcommand {\barr}{\begin{array}}
\newcommand {\earr}{\end{array}}

\newcommand {\N}{{\cal N}}

\newtheorem{definition}{Definition}
\newtheorem{property}[definition]{Property}
\newtheorem{proposition}[definition]{Proposition}
\newtheorem{lemma}[definition]{Lemma}
\newtheorem{theorem}[definition]{Theorem}
\newtheorem{corollary}[definition]{Corollary}
\newtheorem{example}[definition]{Example}
\newtheorem{remark}[definition]{Remark}

\newcommand {\benum} {\begin{enumerate}}
\newcommand {\eenum} {\end{enumerate}}

\newcommand {\bdesc} {\begin{description}}
\newcommand {\edesc} {\end{description}}

\newcommand {\bfig}[2] {\begin{figure}
  \centering
  \includegraphics[width=#2]{#1}}
\newcommand {\brotatefig}[2] {\begin{figure}[htbp]
                        \centerline {
                         \epsfig{figure={#1},clip=,angle=-90,width={#2}}}}

\newcommand {\bfigfirst}[2] {\begin{figure}[h]
                        \centerline {
                        \setlength{\epsfxsize}{#2}
                        \epsffile{#1}}}
\newcommand {\efig}[2]{ \caption{#2}
                        \label{fig:#1}
                        \end{figure}
                        \mymarginpar{fig:#1}}
\newcommand {\erotatefig}[2]{ \caption{#2}
                        \label{fig:#1}
                        \end{figure}
                        \mymarginpar{fig:#1}}
\newcommand {\rfig}[1]{Figure \ref{fig:#1}}

\newcommand {\btab}[1]{
                       \begin{table}
                       \centering
                       \begin{tabular}{#1}}
\newcommand {\etab}[3] {
                       \end{tabular}
                       \caption[#3]{#2}
                       \label{tab:#1}
                       \end{table}
                       \mymarginpar{tab:#1}
                       \vspace{.1in}}

\newcommand {\btabular}[1]{\begin{center}
                       \begin{tabular}{#1}}
\newcommand {\etabular}{\end{tabular}
                       \end{center}}

\newcommand {\bdefin}[1]{\begin{definition}
                      \mymarginpar{def:#1}
                      \label{def:#1} }
\newcommand {\edefin}       {\end{definition}}

\newcommand {\bpro}[1]{\begin{property}
                      \mymarginpar{pro:#1}
                      \label{pro:#1} }
\newcommand {\epro}   {\end{property}}

\newcommand {\bprop}[1]{\begin{proposition}
                      \mymarginpar{prop:#1}
                      \label{prop:#1} }
\newcommand {\eprop}       {\end{proposition}}
\newcommand {\rprop}[1]{Proposition \ref{prop:#1}}

\newcommand {\blem}[1]{\begin{lemma}
                      \mymarginpar{lem:#1}
                      \label{lem:#1} }
\newcommand {\elem}   {\end{lemma}}
\newcommand {\rlem}[1]{Lemma \ref{lem:#1}}

\newcommand {\bthe}[1]{\begin{theorem}
                      \mymarginpar{the:#1}
                      \label{the:#1} }
\newcommand {\ethe}   {\end{theorem}}
\newcommand {\rthe}[1]{Theorem \ref{the:#1}}

\newcommand {\bproof}{\noindent {\bf Proof.} \ }
\newcommand {\eproof} {\hfill \squares \\ \vspace{.3cm}}
\newcommand {\bcor}[1]{\begin{corollary}
                      \mymarginpar{cor:#1}
                      \label{cor:#1} }
\newcommand {\ecor}   {\end{corollary}}

\newcommand {\bax}[1]{\begin{axiom}
                      \mymarginpar{ax:#1}
                      \label{ax:#1} }
\newcommand {\eax}       {\vspace{-.1in} \end{axiom}}

\newcommand {\bex}[2]{\vspace{.1in}
                      \begin{example}
                      \mymarginpar{ex:#1}
                       {\bf #2}
                      \label{ex:#1} }
\newcommand {\eex}       {\end{example} \vspace{.3cm} }

\newcommand {\brem}[1]{\begin{remark}
                      \mymarginpar{rem:#1}
                      \label{rem:#1} \em }
\newcommand {\erem}   {\end{remark}}

\newcommand {\beq}[1]{\mymarginpar{eq:#1}
                      \begin{equation}
                      \label{eq:#1} }

\newcommand {\beqno}[1]{\mymarginpar{eq:#1}
                      \begin{eqnarray}
                      \nonumber}

\newcommand {\eeq}       {\end{equation}}
\newcommand {\eeqno}       { && \end{eqnarray}}
\newcommand {\req}[1]{(\ref{eq:#1})}

\newcommand {\bear}[1]{\mymarginpar{eq:#1}
                       \begin{eqnarray}
                       \label{eq:#1} }

\newcommand {\bearno}[1]{\mymarginpar{eq:#1}
                       \begin{eqnarray}
                       \nonumber}

\newcommand {\eear}{\end{eqnarray}}
\newcommand {\eearno}{\end{eqnarray}}
\newcommand {\bsel}{\left \{ \begin{array}{cl}}
\newcommand {\esel}{\end{array} \right.}

\newcommand {\bmat}[1]{\left [ \begin{array}{#1}}
\newcommand {\emat}{\end{array} \right ]}
\newcommand {\bsec}[2]{\mymarginpar{sec:#2}
                       \section{#1}
                       \label{sec:#2} }

\newcommand {\rsec}[1]{Section \ref{sec:#1}}

\newcommand {\bsubsec}[2]{\mymarginpar{sec:#2}
                       \subsection{#1}
                       \label{sec:#2} }

\def\R{I\kern-0.30em R}
\def\N{I\kern-0.30em N}
\def\P{I\kern-0.30em P}

\newcommand\squares{\vrule height6pt width7pt depth1pt}

\newcommand\qq{q}

\begin{document}

\title{Community Detection in Signed Networks: an Error-Correcting Code Approach}

\author{Cheng-Shang Chang, Duan-Shin Lee, Li-Heng Liou, and  Sheng-Min Lu \\
Institute of Communications Engineering,
National Tsing Hua University \\
Hsinchu 30013, Taiwan, R.O.C. \\
Email:
 cschang@ee.nthu.edu.tw; lds@cs.nthu.edu.tw; dacapo1142@gmail.com; s103064515@m103.nthu.edu.tw
}

\maketitle
\begin{abstract}

In this paper, we consider the community detection problem in signed networks, where there are two types of edges: positive edges (friends) and negative edges (enemies). One renowned theorem of signed networks, known as Harary's theorem, states that structurally balanced signed networks are clusterable. By viewing each cycle in a signed network as a parity-check constraint, we show that the community detection problem in a signed network with two communities is equivalent to the decoding problem for a parity-check code. We also show how one can use two renowned decoding algorithms in error-correcting codes for community detection in signed networks: the bit-flipping algorithm, and the belief propagation algorithm. In addition to these two algorithms,  we also propose a new community detection algorithm, called the Hamming distance algorithm, that performs community detection by finding a codeword that minimizes the Hamming distance.  We compare the performance of these three algorithms by conducting various experiments with known ground truth. Our experimental results show that our Hamming distance algorithm outperforms the other two.
\end{abstract}

\bsec{Introduction}{introduction}
As the advent of on-line social networks, structural analysis of networks becomes an important research topic. In a social network, an edge between two nodes usually represents friendly interactions between these two nodes, and structural analysis of   networks with both directed or undirected edges has been studied extensively in the literature (see e.g., \cite{lancichinetti2009community,newman2010networks,malliaros2013clustering} and references therein).  However, as pointed out in the recent survey paper \cite{tang2015survey}, structural analysis of signed networks has received a lot of attention lately in various areas, including sociology, physics, biology, and computer science. In signed networks, there are two types of edges: positive edges  (friends) and negative edges (enemies). With these two types of edges in signed networks, researchers can better characterize the interactions between two persons in social networks.

 A signed network is called {\em structurally balanced} if every cycle in the network contains an even number of negative edges \cite{cartwright1956structural,newman2010networks}. One of the most renowned theorems of signed networks is Harary's theorem \cite{harary1953notion}.  Harary's theorem  states  that a structurally balanced signed network is clusterable and it can be separated into several communities, where the edges within a community are positive and the edges between two different communities are negative (see the node coloring algorithm in Algorithm 1 for more detailed explanations). But what if a signed network is not structurally balanced? How do we address the community detection problem in a signed network that is not structurally balanced?

 In this paper, we focus on the community detection problem in signed networks that might not be structurally balanced. For such a problem,
it was proposed in  \cite{doreian1996partitioning}  a partition criterion (and an associated partitional algorithm)  that finds a partition of the nodes to minimize the (weighted) sum of the following two types of errors: (i) the number of positive edges between two different communities and (ii) the number of negative edges within a community. Such an intuition for the two types of errors in signed networks  can be  easily incorporated into the notion of modularity \cite{NG04} for community detection in unsigned networks. In particular, it was proposed in \cite{gomez2009analysis} that community detection in signed networks could be formulated as an optimization problem that maximizes the positive modularity and minimizes the negative modularity.
In addition to modularity maximization, a signed Laplacian matrix was proposed in \cite{kunegis2010spectral} and the community detection problem in signed networks was formulated as a normalized cut for such a matrix.

One of the main contributions of this paper is to address the community detection problem in signed networks by using error-correcting codes (ECC).
To the best of our knowledge, this is the first paper that links error-correcting codes to the community detection problem in signed networks.
The fundamental insight of this is that each cycle in a signed network can be viewed  a parity-check constraint in an error-correcting code and every structurally balanced signed network can be viewed as a legitimate codeword. A signed network that is not structurally balanced can then be ``corrected'' back to a structurally balanced sign network by changing a ``minimum'' number of the signs of the edges. In this paper, we show how one can use two renowned decoding algorithms in error-correcting codes for community detection in signed networks with two communities: (i) the bit-flipping algorithm, (ii) the belief propagation algorithm.
 In addition to these two algorithms,  we also propose a new community detection algorithm, called the Hamming distance algorithm, that performs community detection by finding a codeword that minimizes the Hamming distance.  We compare the performance of these three algorithms by conducting various experiments with known ground truth. Our experimental results show that our Hamming distance algorithm outperforms the other two.

The rest of the paper is organized as follows. In \rsec{signed}, we introduce the related backgrounds for signed networks and Harary's theorem. In \rsec{detection}, we establish the links between the community detection problem in signed networks and error-correcting codes. We also illustrate how one can use  two renowned decoding algorithms for community detection in signed networks. Experimental results for these three algorithms are presented in \rsec{exp}. The paper is concluded in \rsec{conclusion}.

\bsec{Signed networks}{signed}

In this paper, we consider community detection in signed networks. A signed network $G_s=(V,E,W)$ consists of a set of nodes $V$, a set of edges $E$, and a function $W$ that maps every edge in $E$ to the two signs $\{+,-\}$. An edge $(u,v)$ with
the sign $+$ is called a positive edge and it is generally used for indicating the {\em friendship} between the two nodes $u$ and $v$. On the other hand, an edge  with the sign $-$ is called a negative edge. A negative edge $(u,v)$ indicates that $u$ and $v$ are enemies and it is better not to cluster these two nodes in the same community. One of the most important results of signed networks is Harary's theorem (see e.g., the book \cite{newman2010networks}). For a signed network, it is said to be {\em structurally balanced} if it contains
only cycles (loops) with even numbers of negative edges. Harary's theorem says that a structurally
balanced signed network is {\em clusterable} in the sense that it can be divided into connected groups
of nodes such that all edges between members of the same group are positive and all
edges between members of different groups are negative. The converse statement for Harary's theorem is in general not true, e.g., a signed network with three nodes connected by three negative edges has a cycle of three negative edges. But it is true for a signed networks with two groups. This is stated in the following proposition.

\bprop{harary}
Consider a signed network that can be divided into {\em two} connected groups
of nodes such that all edges between members of the same group are positive and all
edges between members of different groups are negative. Then every cycle of the signed network
has an even number of negative edges. Thus, the signed network is structurally balanced.
\eprop

\bproof
Image the two groups of nodes as two islands and the negative edges as bridges between these two islands. As there are exactly two groups in the signed network, a cycle that starts from a node in one group must cross the bridges an even number of times in order to get back to where the cycle is started.
\eproof

In view of \rprop{harary}, a structurally
balanced signed network with two groups can be easily detected by the {\em node coloring} algorithm (see e.g., the book \cite{west2001introduction}) that colors all the nodes in two colors, say black and white. It starts from coloring a node with one color (e.g., black) and repeatedly coloring a neighbor of a colored node by the same (resp. the other) color if it is connected by a positive (resp. negative) edge. This is summarized in Algorithm \ref{alg:coloring}.

\begin{algorithm}[t]
\KwIn{A structurally
balanced signed network $G_s=(V,E,W)$.
}
\KwOut{A partition ${\cal P}=\{S_1, S_2\}$.}

\noindent {\bf (1)}  Initially, choose a node $w$ and assign that node to $S_1$.

\noindent {\bf (2)} Traverse the graph by using the breadth-first search (BFS) with node $w$ as the root.

\noindent {\bf (3)} For each neighbor $v$ of a traversed node $u$, assign node $v$ to the set that $u$ is assigned if $(u,v)$ is a positive  edge. Otherwise, assign node $v$ to the other set.

\caption{The Node Coloring Algorithm for a Structural
Balanced Signed Network with Two Groups}
\label{alg:coloring}
\end{algorithm}

\bsec{Community detection}{detection}

In the previous section, we have shown that a structurally
balanced signed network with two groups can be easily detected by the node coloring algorithm in Algorithm \ref{alg:coloring}. But what if the input signed network is not structurally balanced? Our idea is to treat this as an error-correcting code problem and ``correct'' a signed network that is not structurally balanced into another structurally balanced  signed network.

\bsubsec{Parity-check codes}{check}

In this section, we briefly review the parity check codes (see e.g., \cite{blahut1983theory} for more details).
 The Galois field $\mbox{GF}(2)$ defines two operations $\oplus$ (the exclusive-OR operation) and $\cdot$ (the AND operation) on the set $\{0,1\}$.
  These two operations act similarly to the usual addition operation and the usual multiplication operation as they satisfy various algebraic properties, including the associative law, the commutative law and the distributive law. As such, we can add, subtract, multiply and divide in $\mbox{GF}(2)$ as in rational numbers. Specifically, for two binary $m$-vectors ${\bf h}=(h_{1}, h_{2}, \ldots, h_{m})$ and ${\bf w}=(w_1, w_2, \ldots, w_m)$, its inner product is defined as
\beq{check0011a}
{\bf h} \cdot {\bf w}^T=(h_1 \cdot w_1)\oplus (h_2 \cdot w_2) \oplus \ldots \oplus(h_m \cdot w_m),
\eeq
where ${\bf w}^T$ is the transpose of $\bf w$.
The matrix multiplication can also be defined similarly.

Now consider an $\ell \times m$ matrix ${\bf H}=(h_{i,j})$ with all its elements in $\mbox{GF}(2)$ and the set of vectors in the null space of $\bf H$, i.e.,
$\{{\bf w}: {\bf H} \cdot {\bf w}^T={\bf 0}_\ell$\}, where  ${\bf 0}_\ell$ is the zero vector with dimension $\ell$.
The matrix ${\bf H}$ is called the {\em parity-check matrix} and the set of vectors in the null space of $\bf H$ are called the {\em codewords} of the parity-check code with the parity-check matrix ${\bf H}$. To see the intuition behind such an error-correcting code, suppose we transmit a codeword $\bf w$ through an error-prone channel and receive another vector ${\bf w}^\prime$. The vector ${\bf w}^\prime$ may not be in the null space and thus we may be able to correct it by selecting a codeword that is ``closest'' to  ${\bf w}^\prime$. There are a vast amount of papers in the literature addressing the issue of how to ``decode'' the received vector. In this paper, we will consider three renowned decoding algorithms:
the bit-flipping algorithm, the belief propagation algorithm, and the Hamming distance algorithm.

\bsubsec{Cycle basis and parity-check codes}{parity}

In this section, we establish the connections between signed networks and error-correcting codes.
Consider a  graph $G=(V,E)$ with $m$ edges and $n$ nodes. Index these $m$ edges from $1,2,\ldots, m$ and $n$ nodes from $1,2, \ldots, n$. Then every (simple) cycle of the graph can be represented by a binary $m$-vector ${\bf h}=(h_1, h_2, \ldots, h_m)$, where $h_i$ indicates whether the $i^{th}$ edge is in the cycle. Suppose that ${\bf h}$ and ${\bf h}^\prime$ represent two {\em non-disjoint} cycles of the graph. Then it is clear that ${\bf h} \oplus {\bf h}^\prime$ also represents another cycle of the graph, where $\oplus$ is the bit-wise exclusive-OR operation of these two vectors. In fact, any linear combination of cycles (in the field of $\mbox{GF}(2)$) is also a cycle (or a collection of disjoint cycles). Thus, a collection of the maximum number of linearly independent cycles forms a cycle basis. For a connected graph with $n$ nodes and $m$ edges, the maximum number of linearly independent cycles is known to be $m-n+1$ \cite{horton1987polynomial}. Thus, every collection of $m-n+1$ linearly independent cycles forms a cycle basis. One way to find a cycle basis is to first construct a spanning tree of the connected  network with $m$ edges and $n$ nodes. The number of edges in the spanning tree is $n-1$ and there are exactly $m-n+1$ edges that are not in the spanning tree. Adding each of these $m-n+1$ edges to the spanning tree forms a linearly independent cycle.

Suppose for a connected graph $G=(V,E)$ with $m$ edges and $n$ nodes, we have found a cycle basis $\{
{\bf h}_i, i=1,2\ldots, m-n+1\}$, where ${\bf h}_i=(h_{i,1}, h_{i,2}, \ldots, h_{i,m})$. Form the $(m-n+1)\times m$ matrix ${\bf H}=(h_{i,j})$.
Such a  matrix ${\bf H}$ is called a fundamental cycle matrix for the graph $G=(V,E)$.
Now for a signed network $G_s=(V,E,W)$, we can also associate each positive edge with the weight 0 and each negative edge with the weight 1. Then the weights of the $m$ edges can be represented by a binary vector ${\bf w}=(w_1, w_2, \ldots, w_m)$. Such a vector $\bf w$ is called the weight vector of the signed network $G_s=(V,E,W)$.
Clearly, the statement that a cycle represented by ${\bf h}=(h_{1}, h_{2}, \ldots, h_{m})$
contains an even number of negative edges is equivalent to
\beq{check0011}
(h_1 \cdot w_1)\oplus (h_2 \cdot w_2) \oplus \ldots \oplus(h_m \cdot w_m)=0.
\eeq
Writing \req{check0011} in the inner product of two vectors in $\mbox{GF}(2)$ yields
\beq{check0022}
{\bf h} \cdot {\bf w}^T =0,
\eeq
 ${\bf w}^T$ is the transpose of ${\bf w}$.

In the following lemma, we show that a structurally balanced signed network is a codeword of a parity-check code.

\blem{check}
Consider a signed network $G_s=(V,E,W)$.
Let ${\bf H}$ be any fundamental cycle matrix for the graph $G=(V,E)$ and
$\bf w$ be the weight vector of the signed network $G_s=(V,E,W)$.
Then the signed network $G_s=(V,E,W)$  is structurally balanced if and only if
\beq{check1111}
{\bf H} \cdot {\bf w}^T ={\bf 0}_{m-n+1},
\eeq
where the matrix product is in $\mbox{GF}(2)$ and ${\bf 0}_{m-n+1}$ is the $m-n+1$-column vector with all its elements being 0.
\elem

\bproof
($\Rightarrow$)
If the signed network $G_s=(V,E,W)$  is structurally balanced, then every cycle in $G_s$ consists of an even number of negative edges. Note that ${\bf H}$ is a fundamental cycle matrix of $G_s$ with its row vector ${\bf h_i}$ representing the $i^{th}$ cycle. By \req{check0011}, we have ${\bf h_i} \cdot {\bf w}^T =0$, for $1\leqslant i \leqslant m-n+1$. Thus,  ${\bf H} \cdot {\bf w}^T ={\bf 0}_{m-n+1}$.

($\Leftarrow$) Assume ${\bf H} \cdot {\bf w}^T ={\bf 0}_{m-n+1}$. Then ${\bf h_i} \cdot {\bf w}^T =0$, for $1\leqslant i \leqslant m-n+1$. Rewrite it as in
\req{check0011}, i.e.,
$$(h_1 \cdot w_1)\oplus (h_2 \cdot w_2) \oplus \ldots \oplus(h_m \cdot w_m)=0.$$
 It implies that the $i^{th}$ cycle consists of an even number of negative edges for all $i$. Thus, the signed network $G_s=(V,E,W)$  is structurally balanced.
\eproof

In view of \req{check1111}, a fundamental cycle matrix $\bf H$ for a graph can be viewed as the parity-check matrix of a parity-check code.
Also, a vector ${\bf w}$ that satisfies \req{check1111} is a codeword for the  parity-check matrix {\bf H} (or simply a codeword for the graph $G=(V,E)$ in this paper).
In the following lemma, we further show that a two-way partition of a network corresponds to a codeword of a parity-check code.

\blem{codeword}
Consider a two-way partition ${\cal P}=\{S_1, S_2\}$ of the nodes in a graph $G=(V,E)$.
Construct a signed network $G_s(V,E,W)$ by assigning all the edges within the same set to be positive and all
the edges between two sets to be negative. Let ${\bf w}({\cal P})$ be the weight vector of the signed network $G_s=(V,E,W)$.
Then ${\bf w}({\cal P})$ is a codeword for the graph $G=(V,E)$.
\elem

\bproof
As a direct result of Harary's theorem in \rprop{harary}, the signed network $G_s=(V,E,W)$ constructed from the partition ${\cal P}=\{S_1, S_2\}$ is structurally balanced. Also, in the structurally balanced network, the multiplication of a fundamental cycle matrix $\bf H$ and ${\bf w}({\cal P})$ is zero by \rlem{check}. Thus, we conclude that ${\bf w}({\cal P})$ is a codeword for the graph $G=(V,E)$.
\eproof

From \rlem{codeword}, we know every two-way partition corresponds to a codeword. This leads to the following method to generate codewords for a graph $G(V,E)$ from an $n$-binary vector ${\bf x}=(x_1, x_2, \ldots, x_n)$. For the $n$-binary vector, we assign node $i$ the value $x_i$ for $i=1,2, \ldots, n$. Suppose that the two ends of the $j^{th}$ edge are node $u$ and $v$. Then we assign  $w_{j}=x_u \oplus x_v$ for all $j=1,2, \ldots, m$. Note that both the two binary vectors ${\bf 0}_n$ and ${\bf 1}_n$ generate the same codeword, i.e., the zero codeword. Excluding these two binary vectors, we let $S_1=\{i: x_i =0\}$ and $S_2=\{i: x_i=1\}$ and this yields a two-way partition and a codeword ${\bf w}({\cal P})$. We can write this in the following matrix form:
\beq{generator1111}
{\bf w} = {\bf x}\cdot {\bf G},
\eeq
where ${\bf G}=(g_{i,j})$ is $n \times m$ generator matrix with $g_{i,j}=1$ when node $i$ is one end of the $j^{th}$ edge and 0 otherwise.
Such a generator matrix was considered in \cite{abbe2014decoding} for decoding binary node values.

\begin{figure}[h]
\centering
\includegraphics[width=0.25\textwidth]{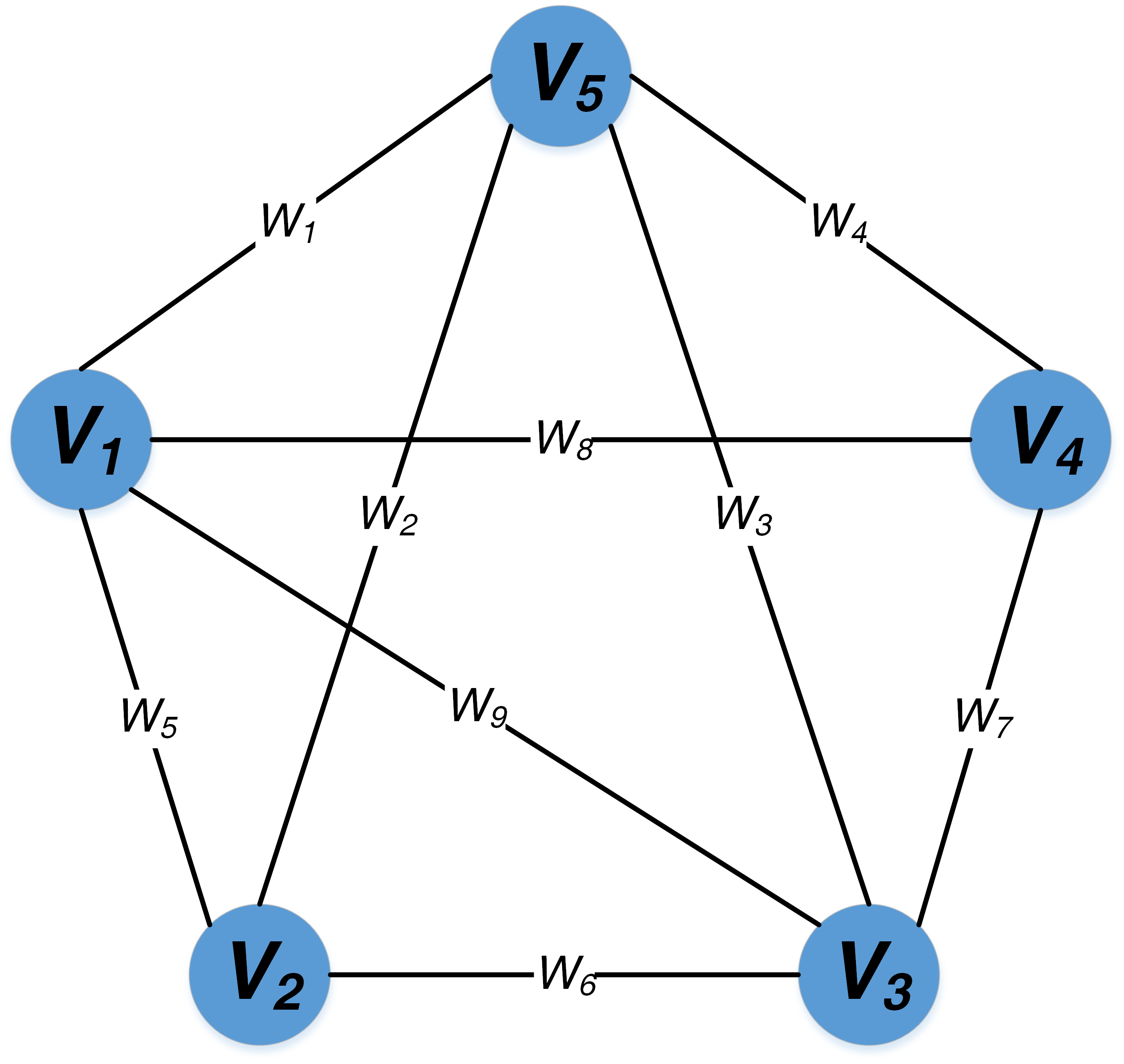}
\caption{A graph with five nodes and nine edges.}
 \label{fig:star}
\end{figure}

To illustrate this, let us consider a graph with five nodes and nine edges
in \rfig{star}.  For this graph, we choose the spanning tree with the edges connected to vertex 5. Such a spanning tree is a star graph and every fundamental cycle contains exactly three edges. For this graph, we have the following parity check matrix
\beq{star1111}
{\bf H}=\left[
\begin{array}{ccccccccc}
1 & 1 & 0 & 0 & 1 & 0 & 0 & 0 & 0 \\
0 & 1 & 1 & 0 & 0 & 1 & 0 & 0 & 0 \\
0 & 0 & 1 & 1 & 0 & 0 & 1 & 0 & 0 \\
1 & 0 & 0 & 1 & 0 & 0 & 0 & 1 & 0 \\
1 & 0 & 1 & 0 & 0 & 0 & 0 & 0 & 1 \\
\end{array}
\right],
\eeq
and the generator matrix
\beq{star2222}
{\bf G}=\left[
\begin{array}{ccccccccc}
1 & 0 & 0 & 0 & 1 & 0 & 0 & 1 & 1 \\
0 & 1 & 0 & 0 & 1 & 1 & 0 & 0 & 0 \\
0 & 0 & 1 & 0 & 0 & 1 & 1 & 0 & 0 \\
0 & 0 & 0 & 1 & 0 & 0 & 1 & 1 & 0 \\
1 & 1 & 1 & 1 & 0 & 0 & 0 & 0 & 0 \\
\end{array}
\right].
\eeq

Our approach for the (two-way) community detection problem for a signed network $G_s(V,E,W)$ is to treat a signed network as a received signal and then decode such a signal by finding the most likely codeword.
In \rsec{flip} and \rsec{belief}, we discuss two
commonly used decoding algorithms for low-density parity-check (LDPC)  codes in the literature (see e.g., \cite{gallager1962low,kou2001low}): (i) the bit-flipping algorithm, and (ii) the belief propagation algorithm. In \rsec{hamming}, we propose our decoding algorithm based on Hamming distance.

\bsubsec{The bit-flipping algorithm}{flip}


For a signed network $G_s(V,E,W)$, let ${\bf w}$ be its weight vector and ${\bf H}$ be a fundamental cycle matrix. As discussed before, we may consider $\bf w$ as the received signal of a parity-check code. To decode such a signal, we first compute the syndrome ${\bf s}=(s_1, s_i, \ldots, s_{m-n+1})$ in $\mbox{GF}(2)$:
\beq{bit01}
s_i=\sum_{j=1}^{m}{h_{ij}w_j}.
\eeq
Rewriting \req{bit01} in the inner product form yields
\beq{bit02}
{\bf s}={\bf H} \cdot {\bf w}^T,
\eeq
where the matrix product is in $\mbox{GF}(2)$. The $i^{th}$ syndrome component, $s_i$, indicates whether the $i^{th}$ cycle in $G=(V,E)$ is structurally balanced. If $s_i$ equals to one, the $i^{th}$ cycle is not structurally balanced. In view of \req{check1111}, all of the parity-check constraints are satisfied if all syndrome components are zero.

With the syndrome, we can then compute  $u_k$ as  follows:
\beq{bit03}
u_k=\sum_{i=1}^{m-n+1}{s_i h_{ik}} ,
\eeq
The physical meaning of $u_k$ is the number of unbalanced cycles that traverse through the $k^{th}$ edge.
 Thus, we can find out the edge that has the largest number of unbalanced cycles and then flip the sign of that edge. Intuitively, such a greedy correction will reduce the number of unbalanced cycles and hopefully it will converge to a codeword (that has no unbalanced cycles).
   Specifically, let $k^*$ be the index of the edge that has the largest number of unbalanced cycles, i.e.,
\beq{bit05}
k^*=\arg\max_{k}{u_k}.
\eeq
We then flip its sign from 1 to 0 or from 0 to 1, i.e.,
\beq{bit04}
w_k\leftarrow 1-w_k .
\eeq
We summarize the bit-flipping algorithm in Algorithm \ref{alg:bitflipping}. However, this algorithm does not guarantee convergence. As pointed out in \cite{kou2001low}, usually there is a design parameter $\delta$ that stops the algorithm when the number of unbalanced cycles is within $\delta$.

\begin{algorithm}[t]
\KwIn{A  signed network $G_s=(V,E,W)$.
}
\KwOut{A partition ${\cal P}=\{S_1, S_2\}$.}

\noindent {\bf (1)} Compute the syndrome {$\bf s$} in \req{bit02}. If all the syndrome components are zero, then stop decoding.

\noindent {\bf (2)} Find the number of unbalanced cycles $u_k$ for each edge $k$ by using \req{bit03}.

\noindent {\bf (3)} Find the edge with maximum $u_k$ and then flip the sign of that edge.

\noindent {\bf (4)} Repeat Steps 1 to 3 until all of the syndrome components are zero or a predefined maximum  number of iterations is reached.

\noindent {\bf (5)} Detect two groups by the {\em node coloring} algorithm in Algorithm \ref{alg:coloring}.

\caption{The Bit-Flipping Algorithm}
\label{alg:bitflipping}
\end{algorithm}


\bsubsec{The belief propagation algorithm}{belief}

The belief propagation algorithm, first introduced by Gallager \cite{gallager1962low}, is a soft decision method. In this paper, we use the computer program written for the sum-product algorithm  in \cite{lee2015iterative} (we thank Prof. H.-C. Lee for providing us the computer program). The sum-product algorithm is to compute the maximum a posteriori probability for each codeword bit.

\begin{figure}[h]
\centering
\includegraphics[width=0.4\textwidth]{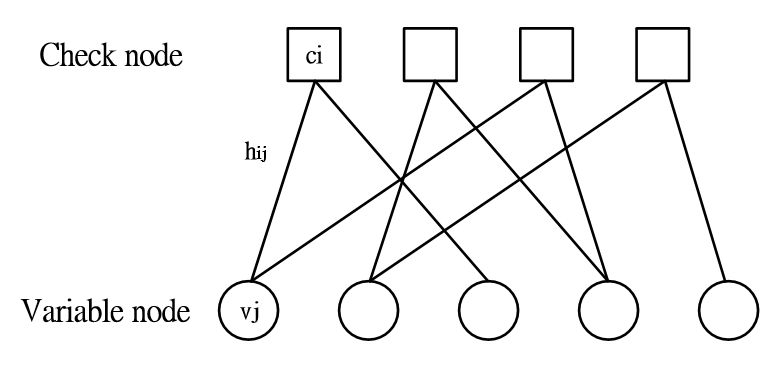}
\caption{The construction of the bipartite graph from $\bf H$}
\label{fig:bipartite}
\end{figure}

For a signed network $G_s(V,E,W)$, let ${\bf w}$ be its weight vector and ${\bf H}=(h_{ij})$ be a fundamental cycle matrix. We can create a bipartite graph as depicted in \rfig{bipartite} with $m$ variable nodes and $m-n+1$ check nodes, where $v_j$ denotes the $j^{th}$ variable node, and $c_i$ denotes the $i^{th}$ check node. If $h_{ij}=1$, there is an edge between variable node $j$ and check node $i$.

Denote by $\mathcal{N}(v_j)$ the neighboring check nodes that connect to variable node $v_j$, i.e., $\mathcal{N}(v_j)=\lbrace c_i:h_{ij}=1\rbrace$.
 Similarly, denote by $\mathcal{N}(c_i)$ the neighboring variable nodes that connect to check node $c_i$, i.e., $\mathcal{N}(c_i)=\lbrace v_j:h_{ij}=1\rbrace$.

For the ease of computation, we  deal with the log-likelihood ratios (LLRs). Using LLRs as messages offers implementation advantages over using probabilities or likelihood ratios, because multiplications are replaced by additions and the normalization step is eliminated. The log-likelihood ratio is shown below:

\beq{BF01}
L(w)=\log(\frac{p(w=0)}{p(w=1)}).
\eeq
If $p(w=0)>p(w=1)$ (resp. $p(w=1)>p(w=0)$) then $L(w)$ is positive (resp. negative). The larger the magnitude of $L(w)$, the higher the probability that $w$ would equal to zero (resp. one).

From the Bayes formula, a posteriori probability is positively correlated to likelihood and a priori probability.
Then, we can compute a posteriori LLR as follows:
\beq{BP_new01}
\log\frac{p(w^{*}=0|w)}{p(w^{*}=1|w)}=\log\frac{p(w|w^{*}=0)}{p(w|w^{*}=1)}+\log\frac{p(w^{*}=0)}{p(w^{*}=1)},
\eeq
where $w^{*}$ is the original sign of the edge and $w$ is the observed sign. On the right-hand side, the first term is the log-likelihood ratio and the second term is a priori LLR. The log-likelihood ratio is called the intrinsic information and can be computed
for the binary symmetric channel as follows:
\beq{BP_new01b}
\log\frac{p(w_j|w_j^{*}=0)}{p(w_j|w_j^{*}=1)}
=\log(\frac{1-p}{p})w_j.
\eeq

We denote a posteriori LLR as the variable-to-check message $L_{v_j \rightarrow c_i}$ that propagates from variable node $v_j$ to check node $c_i$. Such a message can be calculated according to
\beq{BF08}
L_{v_j \rightarrow c_i}=\sum_{c_a\in{\mathcal{N}(v_j)\setminus{c_i}}}{m_{c_a \rightarrow v_j}}+L_{v_j}.
\eeq
where $L_{v_j}$ is the intrinsic LLR of variable node $v_j$.

A priori LLR is denoted as the check-to-variable message $m_{c_i \rightarrow v_j}$ that propagates from check node $c_i$ to variable node $v_j$. Such a message is generated according to
\beq{BF06}
m_{c_i \rightarrow v_j}=2\tanh^{-1}\left(\prod_{v_b\in{N(c_i)\setminus{v_j}}}{\tanh\left(\frac{L_{v_b \rightarrow c_i}}{2}\right)}\right).
\eeq

Thus, we can compute the LLR value of variable node $v_j$ by using
\beq{BF09}
L(v_j)=\sum_{c_i\in{\mathcal{N}(v_j)}}{m_{c_i \rightarrow v_j}}+L_{v_j}.
\eeq

To estimate the value of the $j^{th}$ variable node $v_j$, we can make a hard decision according to
\beq{BF11}
\hat w_j = \left\{
\begin{array}{rr}
0, & \mbox{if $L(v_j) \geq 0$} \\
1, & \mbox{if $L(v_j) < 0$}
\end{array} \right. .
\eeq

With a hard decision value $\bf\hat w$, we can compute the syndrome $s$ by using \req{bit02}. If all of the syndrome components are zero, then $\bf\hat w$ is a codeword for the graph $G=(V,E)$. Conversely, If $\bf\hat w$ does not satisfy all the parity-check constraints, we continue the iteration until all parity-check constraints are satisfied or a predefined maximum  number of iterations is reached. We summarize the belief propagation algorithm in Algorithm \ref{alg:belief}.

\begin{algorithm}[t]
\KwIn{A  signed network $G_s=(V,E,W)$.
}
\KwOut{A partition ${\cal P}=\{S_1, S_2\}$.}

\noindent {\bf (1)} Initially, set $L_{v_j}=w_j\log(\frac{p}{1-p})$.

\noindent {\bf (2)} For each check node $c_i$, update $m_{c_i \rightarrow v_j}$ by using \req{BF06}.

\noindent {\bf (3)} For each variable node $v_j$, update $L_{v_j \rightarrow c_i}$ by using \req{BF08}.
and update $L_{v_j}$ by using \req{BF09}

\noindent {\bf (4)} Compute the decision value $\hat w_j$ by using \req{BF11}.

\noindent {\bf (5)} Repeat Steps 1 to 4 until all of the parity-check constraints are satisfied or a predefined maximum  number of iterations is reached.

\noindent {\bf (6)} Detect two groups by the {\em node coloring} algorithm in Algorithm \ref{alg:coloring}.

\caption{The Belief Propagation Algorithm}
\label{alg:belief}
\end{algorithm}

\bsubsec{The Hamming distance algorithm}{hamming}

The Hamming distance between two binary vectors is the number of different bits between these two binary vectors.
For a signed network $G_s(V,E,W)$, let ${\bf w}$ be its weight vector and ${\bf w}({\cal P})$ be the weight vector associated with the partition ${\cal P}=\{S_1, S_2\}$ of the nodes in the graph $G=(V,E)$. The Hamming distance algorithm aims to find a codeword that has the minimum Hamming distance to the received signal. In the following lemma, we first show how to compute the Hamming distance by computing the number of positive (resp. negative) edges between two sets.

\blem{Hammdis}
The Hamming distance between
${\bf w}$ and ${\bf w}({\cal P})$, denoted by $d({\cal P})$, can be computed as follows:
\bear{hamm1111}
d({\cal P})&=&\frac{1}{2} \Big (N^-(S_1,S_1)+N^+(S_1,S_2)\nonumber\\
&&\quad+N^+(S_2,S_1)+N^-(S_2,S_2) \Big),
\eear
where
\beq{hamm2222}
N^-(S_i,S_j)=\sum_{u \in S_i, v \in S_j}{\bf 1}\{w(u,v)=1\}
\eeq
is the number of negative edges from the set $S_i$ to the set $S_j$,
and
\beq{hamm3333}
N^+(S_i,S_j)=\sum_{u \in S_i, v \in S_j}{\bf 1}\{w(u,v)=0\}
\eeq
is the number of positive edges from the set $S_i$ to the set $S_j$.
\elem

\bproof
Note that
\bear{hamm1155}
d({\cal P})&=&\sum_{i=1}^m{\bf 1}\{w_i=1, w({\cal P})_i=0\} \nonumber\\
&&\quad +\sum_{i=1}^m{\bf 1}\{w_i=0, w({\cal P})_i=1\}.
\eear
For the $i^{th}$ edge, we know that $w({\cal P})_i=0$ if both ends of the $i^{th}$ edge belong to the same set.
Thus,
\bearn
&&\sum_{i=1}^m{\bf 1}\{w_i=1, w({\cal P})_i=0\}\\
&&=\frac{1}{2}N^-(S_1,S_1)+\frac{1}{2}N^-(S_2,S_2).
\eearn
On other hand, $w({\cal P})_i=1$ if one end of the $i^{th}$ edge belongs to one set and the other end belongs to the other set.
Thus,
\bearn
&&\sum_{i=1}^m{\bf 1}\{w_i=0, w({\cal P})_i=1\}\\
&&=\frac{1}{2}N^+(S_1,S_2)+\frac{1}{2}N^+(S_2,S_1).
\eearn
\eproof

In general, there is no efficient algorithm to find the optimal ${\bf w}({\cal P})$
that minimizes $d({\cal P})$. Here we propose a heuristic algorithm in Algorithm \ref{alg:partitional} that finds a local optimum.
Such an algorithm is related to the partitional algorithm in \cite{doreian1996partitioning}.
 Algorithm \ref{alg:partitional} can be clearly described by defining the correlation measure between a node $v$ and a set $S$ as the difference of the number of positive edges and the number of negative edges from $v$ to $S$, i.e.,
\beq{hamm4444}
\qq(v,S)=N^+(\{v\},S)-N^-(\{v\},S).
\eeq
 With this correlation measure, Algorithm \ref{alg:partitional} is simply a local search algorithm that iteratively assigns each node to the most correlated set.

\begin{algorithm}[t]
\KwIn{A  signed network $G_s=(V,E,W)$.
}
\KwOut{A partition ${\cal P}=\{S_1, S_2\}$.}

\noindent {\bf (0)} Initially, choose arbitrarily two disjoint nonempty sets $S_1$ and $S_2$ as a partition of the $n$ nodes in $G=(V,E)$.

\noindent {\bf (1)} \For{$v=1, 2, \ldots, n$}{

\noindent
{Compute the correlation measures
$\qq(v, S_1)$ and $\qq(v, S_2)$ in \req{hamm4444}.}

\noindent If the two correlation measures are the same, node $v$ remains in the original set.
 Otherwise, assign node $v$ to the set with a larger correlation measure.}

\noindent {\bf (2)} Repeat from Step 1 until there is no further change.
\caption{The Hamming Distance Algorithm}
\label{alg:partitional}
\end{algorithm}

Unlike the bit-flipping algorithm and the belief propagation algorithm, we show in the following theorem that the Hamming distance algorithm in \ref{alg:partitional} is guaranteed to converge within a finite number of steps.
The proof of \rthe{partitional} is given in Appendix A.

\bthe{partitional}
In  Algorithm \ref{alg:partitional}, the Hamming distance is non-increasing when there is a change, i.e.,  a node is moved from one set to another. Thus, the algorithm converges to a local minimum of the Hamming distance in a finite number of steps.
\ethe

\bsec{Experimental results}{exp}

\begin{figure*}[tb]
    \begin{center}
    \begin{tabular}{p{0.3\textwidth}p{0.3\textwidth}p{0.3\textwidth}}
      \includegraphics[width=0.3\textwidth]{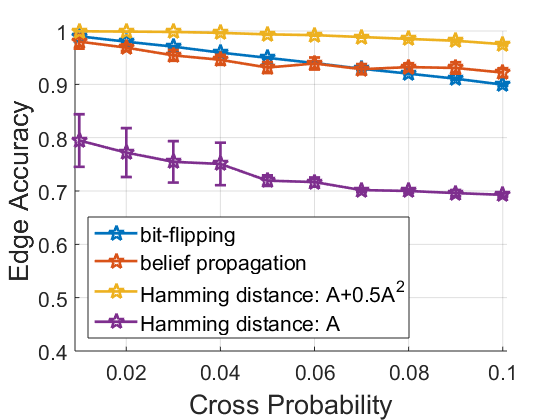} &
      \includegraphics[width=0.3\textwidth]{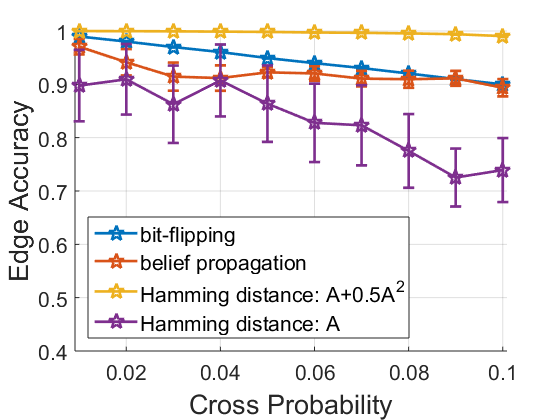} &
      \includegraphics[width=0.3\textwidth]{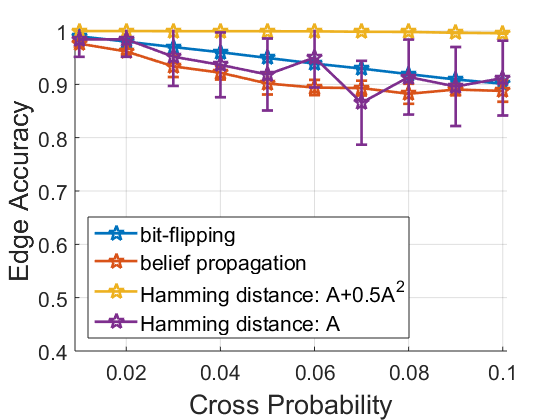}\\
      (a) average degree=6 & (b) average degree=8 & (c) average degree=10 \\ \\
    \end{tabular}
    \caption{Community detection with two communities.}
    \label{fig:ECC}
  \end{center}
\end{figure*}

\bsubsec{Community detection with two communities}{setup}

In this section, we conduct experiments for these ECC algorithms by using the stochastic block model. The stochastic block model, commonly used for benchmarking community detection algorithms, is a generalization of the Erd\"os-R\'enyi model. In the stochastic block model with $n$ nodes and two blocks, the two blocks are equally sized with $n/2$ nodes. The parameter $p_{in}$ is the probability that there is a positive edge between two nodes within the same block and $p_{out}$ is the probability that there is a negative edge between two nodes in two different blocks. All edges are generated independently according to $p_{in}$ and $p_{out}$. Let $c_{in}=np_{in}$ and $c_{out}=np_{out}$. Clearly, such a construction generates a structurally balanced signed network with two ground truth communities.

To generate  a  signed network that is not structurally balanced,
 we randomly flip the sign of an edge in the stochastic block model with the crossover probability $p$. Clearly, if $p$ is small, the signed network is not too far from the original  structurally balanced signed network and it is more likely that we can recover the original signed network. The method of random bit-flipping corresponds to the  binary symmetric channel in a communication system.

In our experiments, the total number of nodes in the stochastic block model is 2000 with 1000 nodes in each block. The parameter $c$ denotes the average degree of a node that is set to be 6, 8, and 10. The value of $c_{in}-c_{out}$ is set to be 5.
The crossover probability $p$ is in the range from 0.01 to 0.1 with a common step of 0.01. We generate 20 graphs for each $p$ and $c$. We remove isolated nodes, and thus the exact numbers of nodes in the experiments might be less than 2000. We show the experimental results with each point averaged over 20 random graphs. The error bars represent the 95\% confident intervals.

For the bit-flipping algorithm (resp. belief propagation algorithm), the maximum number of iterations is set to be 20 (resp. 100).
 To test the Hamming distance algorithm, we conduct our experiments with two adjacency matrices: one with the original adjacency matrix $A$  and  the other with the two-step adjacency matrix
\beq{BF11b}
\widehat{A}=A+0.5A^2.
\eeq
The intuition of using the two-step adjacency matrix is that it allows us to ``see'' more than one step relationship between two nodes.
According to Heider's balance theory \cite{heider1946attitudes}, ``an enemy of my enemy is likely to be my friend'' and ``a friend of my friend is likely to be my friend.'' The two-step adjacency matrix in \req{BF11b}  somehow predicts the two-step relationship between two nodes and thus makes the signed network more dense and complete.

We show our experimental results in \rfig{ECC}.
The Hamming distance algorithm with the two-step adjacency matrix
has the largest edge accuracy (defined as the ratio of the number of accurately decoded edges to the total number of edges) for the entire range of the cross probability.
It seems that the corrupted edges are all corrected by using the Hamming distance algorithm with the two-step adjacency matrix. As explained before, the reason that the Hamming distance algorithm with the two-step adjacency matrix is better than that with the original adjacency matrix is because the former can ``see'' more than one step relationship between two nodes.
From \rfig{ECC}, both the bit-flipping algorithm and the belief propagation algorithm do not perform well for community detection in signed networks. One possible explanation is that signed networks in general do not correspond to good error-correcting codes. This is because there might exist the girth-4 problem (see e.g., \cite{zhang2003design,fan2008design}), i.e.,
 the bipartite graph constructed from a fundamental cycle matrix might have cycles of length four and   messages are likely to be trapped in short cycles. The girth-4 problem can be avoided  by using some known methods in \cite{zhang2003design,fan2008design} by constructing a new parity-check matrix. However, these methods cannot be used here as the parity-check  matrix $\bf H$ is constructed from the spanning tree of a random graph.

From our experimental results, it seems that the performance of the bit-flipping algorithm is slightly better than that of the belief propagation algorithm.
Also, increasing the degree in the stochastic block model seems to have a positive effect on the Hamming distance algorithm with the original adjacency matrix. This might be due to the fact that the tested signed networks are more dense. However, increasing the degree in the stochastic block model seems to result in little improvement
for the other two algorithms.

\bsubsec{Community detection with a real dataset}{real}

\begin{figure}[h]
\centering
\includegraphics[width=0.4\textwidth]{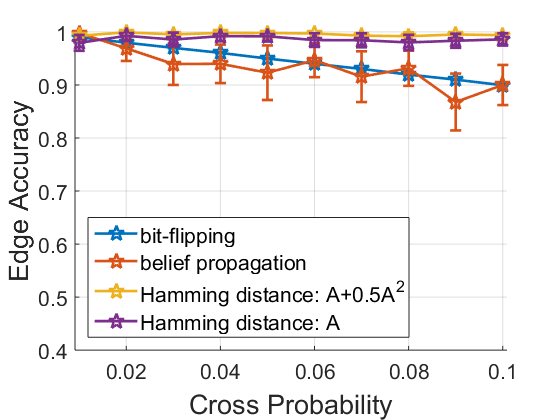}
\caption{Community detection for the political blogs dataset \cite{adamic2005political}.}
\label{fig:real}
\end{figure}

In this section, we report the experimental results for the three ECC algorithms based on the political blogs dataset \cite{adamic2005political}. The dataset contains a directed citation network which is based on a single day snapshot of 1494 political blogs. Each link is established if there is hyperlink from one blog to another blog. For each node, there is an attribute indicating the political orientation (i.e., conservative or liberal) of the blog. To generate a signed network, we convert the network into an undirected graph and label the edges within the same community (resp. between two communities) to +1 (resp. -1). Then, we flip the signs of edges randomly as we did in \rsec{setup}. To ensure the connectivity, we remove all the isolated nodes. By doing so, the number of nodes is down to 1222 and the average degree is about 31. The sizes of these two communities are 586 and 636, respectively. All the remaining setup and parameters for each algorithm are the same as in \rsec{setup}. The main difference between the political blogs dataset and the stochastic block model is that the political blogs dataset has different edge densities in the two communities. The conservative community has a denser edge density than that of the liberal community. The experimental results in \rfig{real} are consistent with our early findings in \rsec{setup} for the stochastic block models.

\bsec{Conclusion}{conclusion}

  In this paper, we considered the community detection problem in signed networks. By using Harary's theorem,
we showed that the community detection problem in a signed network with two communities is equivalent to the decoding problem for a parity-check code. To the best of our knowledge, this is the first result that links error-correcting codes to the community detection problem in signed networks. We also showed how the bit-flipping algorithm, the belief propagation algorithm, and the Hamming distance algorithm can be used for community detection in signed networks.  We compared the performance of these three algorithms by conducting various experiments with known ground truth. Our experimental results show that the Hamming distance algorithm outperforms the other two.
One possible explanation for this is  that signed networks are in general not good error-correcting codes as there might be short cycles in such networks. As such, the bit-flipping algorithm and the belief
propagation algorithm  do not work well for community detection in signed networks.

\section*{Appendix A}

\setcounter{section}{1}

In this section, we prove \rthe{partitional}.

It suffices to show that if node $v$ is in a set $S_1$  and $\qq(v, S_2) > \qq(v, S_1)$, then move node $v$ from $S_1$ to $S_2$ decreases the Hamming distance.
Also let ${\cal P}$ (resp. ${\cal P}^\prime$) be the partition  before (resp. after) the change.
Note that $N^+(\cdot,\cdot)$ and $N^-(\cdot,\cdot)$ are symmetric, $N^+(v,v)=0$, and $N^-(v,v)=0$. It follows from \req{hamm1111} that
\bearn
&&d({\cal P}^\prime) -d({\cal P})\nonumber\\
&&=\frac{1}{2} \Big (N^-(S_1\backslash \{v\},S_1\backslash \{v\})+N^+(S_1\backslash \{v\},S_2\cup \{v\})\nonumber\\
&&\quad+N^+(S_2\cup \{v\},S_1\backslash \{v\})+N^-(S_2\cup \{v\},S_2\cup \{v\}) \Big)\nonumber\\
&&-\frac{1}{2} \Big (N^-(S_1,S_1)+N^+(S_1,S_2)\nonumber\\
&&\quad+N^+(S_2,S_1)+N^-(S_2,S_2) \Big).
\eearn
Note that
$$N^-(S_1\backslash \{v\},S_1\backslash \{v\})-N^-(S_1,S_1)=-2N^-(\{v\},S_1),
$$
$$
N^-(S_2\cup \{v\},S_2\cup \{v\})-N^-(S_2,S_2) =2N^-(\{v\},S_2),
$$
and
\bearn
&&N^+(S_1\backslash \{v\},S_2\cup \{v\})-N^+(S_1,S_2)\nonumber\\
&&=N^+(S_1\backslash \{v\},\{v\})+N^+(S_1\backslash \{v\},S_2) \nonumber\\
&&\quad-N^+(S_1\backslash \{v\},S_2)-N^+(\{v\},S_2) \nonumber \\
&&=N^+(S_1\backslash \{v\},\{v\})-N^+(\{v\},S_2) \nonumber \\
&&=N^+(S_1,\{v\})-N^+(\{v\},S_2)\nonumber\\
&&=N^+(\{v\},S_1)-N^+(\{v\},S_2) .
\eearn
Thus,
\bearn
&&d({\cal P}^\prime) -d({\cal P})\nonumber\\
&&=N^-(\{v\},S_2)-N^-(\{v\},S_1) \nonumber\\
&&\quad +N^+(\{v\},S_1)-N^+(\{v\},S_2) \nonumber\\
&&=\qq(v,S_1)-\qq(v,S_2) <0
.
\eearn
As the Hamming distance is non-increasing after a change of the partition, there is no loop in the algorithm.
Since the number of partitions is finite, the algorithm thus converges in a finite number of steps.



\end{document}